# Paramagnetic Collective Electronic Mode and Low Temperature Hybrid Modes in the Far Infrared Dynamics of Orthorhombic $NdMnO_3$

Néstor E. Massa*.[1] Leire del Campo,[2] Domingos De Sousa Meneses,[2] Patrick Echegut,[2] María Jesús Martínez-Lope,[3] and José Antonio Alonso[3]

[1] Laboratorio Nacional de Investigación y Servicios en Espectroscopía Óptica-Centro CEQUINOR, Universidad Nacional de La Plata, C.C. 962, 1900 La Plata, Argentina.

[2] CNRS, CEMHTI UPR3079, Univ. Orléans, F-45071 Orléans, France

[3] Instituto de Ciencia de Materiales de Madrid, CSIC, Cantoblanco, E-28049 Madrid, Spain.

•e-mail: neemmassa@gmail.com

# ABSTRACT

We report on far- and mid-infrared reflectivity of $NdMnO_3$ from 4 K to 300K. Two main features are distinguished in the infrared spectra: active phonons in agreement with the expected for orthorhombic $D_{2h}^{16}$-Pbnm (Z=4) space group remaining constant down to 4 K and a well-defined collective excitation in the THz region due to eg electrons in a d-orbital fluctuating environment. We trace its origin to the $NdMnO_3$ high temperature orbital disordered intermediate phase not being totally dynamically quenched at lower temperatures. This results in minute orbital misalignments that translate in randomize non-static eg electrons within orbitals yielding a room temperature collective excitation. Below $T_N \sim 78$ K, electrons gradually localize inducing long-range magnetic order as the THz band condenses into two modes that emerge pinned to the A-type antiferromagnetic order. They harden simultaneously down to 4 K obeying power laws with $T_N$ as the critical temperature and exponents $\beta \sim 0.25$ and $\beta \sim 0.53$, as for a tri-critical point and Landau magnetic ordering, respectively. At 4K they match known zone center spin wave modes. The power law dependence is concomitant with a second order transition in which spin modes modulate orbital instabilities in a magnetoelectric hybridized orbital/charge/spin/lattice scenario. We also found that phonon profiles also undergo strong changes at $T_N \sim 78$ K due to magnetoelasticity.

.

# INTRODUCTION

Highly correlated oxides are materials in which charge, spin, orbital, and lattice are strongly coupled degrees of freedom that are natural ground for a myriad of interesting physical properties. From the onset, primary candidates were magnetic active 3d ion oxides or close related compounds found by Rare Earth substitution. [1] That is, compounds with technological appeal that have as main attributes either strong electric or magnetic fields sharing the same phase, as in multiferroics, in which the magnetic order is intimate coupled to ferroelectric polarization. [2] These new materials that may lack of long-range order either ferroelectric or magnetic undergo colossal increments of intrinsic properties under applied external fields. They support orbital fluctuations that add to the anharmonic lattice polarizability materializing spin-charge-orbital-lattice couplings in a complex magnetoelectric interplay.[3] The polarizable environment is produced by oxygens in octahedra or tetrahedra subunits of perovskite [4] or hexagonal lattices [5] where incipient correlations of electric dipole variations are associated to $Mn^{3+}$ magnetoelastic induced deformations. $NdMnO_3$ is one of those relevant ionic materials that, at difference of the studied multiferroic family of hexagonal manganites with R=Sc, Y, Ho, Er, Tm, Yb, and Lu, much less attention has been paid. It belongs to the group with rare earth larger ionic radius that is insulator due to the absence of mixed valences states [4] and that it is parent to compounds in which colossal magnetoresistance has been reported.[6] Nd occupies an intermediate place between La and Eu, increasing the topological metastability of the perovskite lattice.[1] Its inclusion decreases the Mn-O-Mn angle in RMnO$_3$ (R=rare earth) yielding a room temperature O'-orthorhombic phase that belongs to the space group $D_{2h}^{16}$-P*bnm*. Below $T_N$~78 K the magnetic phase of NdMnO$_3$ is characterized by ferromagnetic alignment of the Mn moments in the *ab* plane. The 2-fold degenerated $e_g$ orbital breaks down and stabilizes the A–type antiferromagnetic Mn ordering. The departure of the $e_g$ orbital degeneracy (which is related to the $t_{2g}^3 e_{1g}$ of the Mn$^{3+}$ electron configuration) also lowers the Nd$^{3+}$ site point of the centrosymmetric in ideal perovskite sites allowing crystal field excitations being active in the infrared.[7]

Neutron scattering measurements show that anomalies in thermal expansion [8] and specific heat [9] at $T_N$, which have been assigned to anisotropic contributions of the rare earths ions, are related to magnetoelasticity triggering step-like discontinuities in all lattice parameters, being particularly strong and proportional to the Mn magnetic moment along the *b* axis. [10]

At $T_{N(Nd)}$~ 12-20 K, Nd$^{3+}$ orders magnetically along *c* at the same time that Mn$^{3+}$ develops a magnetic component along this direction,[10,11] i.e., in NdMnO$_3$ there are Mn-Mn, Mn-Nd, and Nd-Nd exchange interactions.

On the other hand, our measurements in the high temperature regime of NdMnO$_3$ show an orbital disordered background in which $e_g$ electrons are in fluctuating d-orbitals. Far infrared emissivity yields evidence of this rarefied environment suggesting that the electronic induced mechanisms for colossal magnetoresistance or polar ordering involves orbital/charge and/or spin fluctuations.[12]

Here, we address the infrared properties of the lower temperature O' phase that prevails on cooling below the orbital disordered interval between 800 K and 1200 K. In this phase the octahedral JT distortion is found alternating an staggered pattern of $d_{3x^2-r^2}$ and $d_{3y^2-r^2}$ orbitals in the **_ab_** plane that repeats itself along the *c* axis. The allowed $Q_2$ and $Q_3$ JT distortions compete with octahedral rotations yielding to cooperative effects with marked sublattice deformations of the $GdFeO_3$-type structure basal plane. [4,13]

We found that the number of infrared active vibrational bands is in agreement with space group predictions for orthorhombic $D_{2h}^{16}$-P*bnm.* (Z=4) space group. We also detect a smooth well-defined and rather intense band in the THz region below 100 cm$^{-1}$. We associate its activity to eg electrons in a *d*-orbital fluctuating environment. Remarkably, it condenses at T~80 K into two bands that emerge pinned to the A-type antiferromagnetic order. They obey power laws which are concomitant to a second order phase transition driven by a spin-orbital hybridized instability, with the antiferromagnetic transition as the critical temperature, and exponents $\beta_{Ph} = 0.249 \pm ^{0.002}_{0.006}$ and $\beta_{Sp} = 0.530 \pm ^{0.001}_{0.001}$. At 4 K they are found at energies matching zone center spin waves modes of A-type $PrMnO_3$ and E-type $YMnO_3$ measured by inelastic neutron scattering [14, 15].

# EXPERIMENTAL DETAILS

Polished high quality samples in the shape of 10 mm diameter $NdMnO_3$ pellets from polycrystalline powders were prepared by soft chemistry. The crystal structure at ambient temperature, refined from X-ray diffraction data, corresponds to the conventional orthorhombic space group $D_{2h}^{16}$-*Pbnm*. [4, 12]

Temperature dependent medium (MIR), and far infrared (FIR) near normal reflectivity from 4 K to 300 K was measured between 10 cm$^{-1}$-5000 cm$^{-1}$ with a FT-IR Bruker 113v interferometer at 2 cm$^{-1}$ resolution with samples mounted on a cold finger of a homemade cryostat. Supporting measurements have also been done between 2 cm$^{-1}$ and 60 cm$^{-1}$ in a Bruker 66v/S interferometer at the IRIS-Infrared beamline. of the Berlin Electron Storage Ring (BESSYII-Helmholtz Zentrum Berlin für Materialien und Energie. GmbH). A liquid He cooled bolometer and a deuterated triglycine sulfate (DTGS) detector were employed to completely cover the spectral range of interest.

We used a gold mirror as 100% reference as alternative to gold in-situ evaporation. We found that the agreement of same temperature emissivity and reflectivity spectra makes possible the use of the same sample with both techniques without altering its surface. [12] To avoid the effects of interference fringes due to the semi-transparency of NdMnO3 in the THz region we only used 2 mm or thicker samples being the acceptance criteria their no detection on the band profiles.

Phonon frequencies were computed using a standard multioscillator dielectric simulation. [16]. The dielectric function, $\varepsilon(\omega)$, given by

$$\varepsilon(\omega) = \varepsilon_1(\omega) - i\varepsilon_2(\omega) = \varepsilon_\infty \prod_j \frac{(\omega_{jLO}^2 - \omega^2 + i\gamma_{jLO}\omega)}{(\omega_{jTO}^2 - \omega^2 + i\gamma_{jTO}\omega)}, \quad (1)$$

where $\varepsilon_\infty$ is the high frequency dielectric constant taking into account electronic contributions; $\omega_{jTO}$ and $\omega_{jLO}$, are the transverse and longitudinal optical mode frequencies and $\gamma_{jTO}$ and $\gamma_{jLO}$ their respective damping. Then, the real ($\varepsilon_1(\omega)$) and imaginary ($\varepsilon_2(\omega)$) part of the dielectric function (complex permitivity, $\varepsilon^*(\omega)$) are estimated from fitting [17] the data to the reflectivity R given by

$$R(\omega) = \left| \frac{\sqrt{\varepsilon^*(\omega)} - 1}{\sqrt{\varepsilon^*(\omega)} + 1} \right|^2. \quad (2)$$

It is pertinent to also mention, before closing this section, that the left hand side of eq. (1) is, in fact, an approximation strictly valid for magnetically inert materials. Its use implies that the unknown frequency dependent magnetoeletric coupling constant, α(ω), is set to zero.[18] Our measurements suggest that this coupling constant may be magnon-vibrational mode dependent.

# RESULTS AND DISCUSSION

Figs. 1 shows the far infrared reflectivity spectra of $NdMnO_3$ from 300 K to 4 K. Two main features may be distinguished in the overall spectra: regular bands between 100 cm$^{-1}$ and 800 cm$^{-1}$ assigned to phonons and a broad band at THz frequencies locking below $T_N$ into two soft modes. We will address these two points separately in the following sections.

### *i)  Phonon activity*

A multioscillator fit (eq. 1) of the spectrum at 4 K results in 25 active phonon out of the 25 predicted for the space group Pbnm ($D^{16}_{2h}$-Z=4 ) [19]

$$\Gamma_{IR}(O´) = 9B_{1u} + 7B_{2u} + 9B_{3u} \qquad (3)$$

The results for this fit at 4 K is shown in Fig. 2. Table I shows that the number of modes remains constant in the complete low temperature range.

The low temperature side band at ~235 cm$^{-1}$ and a band split in the stretching region centered at ~590 cm$^{-1}$ emerge as weak shoulders splitting from stronger phonon bands. (Figs. 1 and 2, Table I). They may hint JT lattice distortion enhancements contributing to local structural modifications due to the magnetoelastic volume contraction at ~78 K. At this respect is worth noticing that in the context of orbital physics Saitoh et al [20] pointed that similar shoulder behavior, but from Mn-O stretching bond in LaMnO$_3$, were bound to appear as due to a zone boundary phonon folded back to k≈0 in the orbital ordered state. This kind of mode, that was observed by us in RMn$_2$O$_5$ [21] and reported anomalous in TbMn$_2$O$_5$ [22], does not seem to have a direct counterpart in NdMnO$_3$.

We also found a strong relative change in damping of lattice vibrations at ~190 cm$^{-1}$ well above T$_N$, between 120 K and 4 K (Fig. 3), and frequency deviations from purely anharmonic behavior inferring coupling between external phonons and magnetic ordering starting in the paramagnetic phase. We associated this to magnetoelasticity reported particularly strong along the *b* direction parallel to the Mn magnetic moments.[8]

Changes in the profile of the bands involving antisymmetric stretching vibrational nmodes at ~395 cm$^{-1}$ (Fig. 4) point to strong spin-phonon interactions starting at slightly higher temperatures than ~12-20 K, the onset of the Nd$^{3+}$ and Mn$^{3+}$ magnetic rearrangements along de *c* axis. [11] This change in oscillator strength is similar to the one found in infrared and Raman spectra of RMn$_2$O$_5$ (R=Rare Earth, Bi)[20, 23, 24] and correlates well with those found in NdMnO$_3$ symmetric and antisymmetric Raman active phonon bandwidths supporting the overall view for phonon modulation of the superexchange integral. [25,26.27]

On the other hand, as it is shown in figure 5 for the symmetric stretching breathing band, no new phonon features are detected indicating a static lattice distortion with changing symmetry across $T_N$.

### *ii) THz collective mode and soft bands*

A broad smooth band associated to a collective instability is detected in the paramagnetic phase of $NdMnO_3$ at energies lower than phonon frequencies. Although one might correlate this to the primordial idea for charge density waves put forward by Fröhlich [28] on fluctuating electronic cloud response to lattice deformations [29, 30, 31, 32], in our case, we interpret that feature as consequence of lattice-orbital distortions taking place at higher temperatures. At about 1200 K $NdMnO_3$ passes on cooling from the cubic P*m-3m* (pseudocubic orthorhombic P*mcm*) phase to the room temperature cooperative orthorhombic $D_{2h}^{16}$-P*bnm* structure. [12] within an interval of ~300 K where lattice sites with (O′) or without (O), Jahn-Teller (JT) distorted octahedras coexist in an intermediate orbital disordered phase [33]

The JT distortion in orbitals, where electrons reside, has a definitive contribution to the lattice topology built cooperatively with uniformly distorted octahedras in an perturbed orbital-lattice framework with strong coupling of oxygen vibrations [12], -its macroscopic polarization- and the eg quadrupole. This last is the prevalent view known from long-range X-ray diffraction of the $NdMnO_3$ ambient O' orthorhombic phase. It, however, omits possible small angle random orbital deviations that would go along as mostly undetected fluctuations in which asymmetries eg electrons lay in close related orbital states. In the specific case of NdMnO$_3$, an insulating manganite, the net electric dipole of the collective excitation may be traced to loosely bond e$_g$ electrons in a d-orbital fluctuating environment outcome of the orthorhombic spontaneous perovskite strain distortion. It is compromised by changes in the octahedral tilting and O-Mn-O angle. As the global instability index shows, an increment in the perovskite instability points to a strained lattice structure for NdMnO$_3$ [4]. Those electrons contribute to the THz band in a temperature driven scenario that on cooling toward T$_N$ will show increasing charge and magnetic short-range correlations. Orbital fluctuations will slow down due to exchange adding to Coulomb electron interactions. In other words, localizing electrons, magnetically disordered, in fluctuating orbitals correlate via Coulomb interactions leading to the ambient unstructured broad band. Oscillating electrical dipoles condense at T$_N$ into two smooth soft infrared active bands that harden continuously as the long range magnetic order sets in (Fig. 6). The now magnetically tangled electrons have their motion prevailing over the paramagnetic spin entropy having this as the leading factor inducing gradual magnetic ordering. [34] At 4 K, long-range ferromagnetism is well established in the _ab_.

No long-range structural lattice changes are found modulating magnetic interactions, but interactions, mainly expected in the **_ab_** plane, clearly affect about $T_N$ phonon profiles as result of the competition among orbital-charge-spin-lattice couplings (Figs 3-5).

Those two soft modes are sharp, well defined, and grow continuously in intensity and definition as the temperature is lowered as expected for a displacive second order transition, here, with NdMnO$_3$ passing from paramagnetic to the onset of long range ***ab*** plane ferromagnetic order.

A continuous second order phase transition is characterized by critical exponents that in our case are found adjusting a power law to the experimental data using

$$\omega_{soft} = A \cdot (T_{Cr} - T)^{\beta} \tag{4}$$

being A a constant, and $T_{Cr}$ an effective critical temperature

Peaking at ~40 cm$^{-1}$, our higher frequency band departs from the profile expected for a pure electric dipole. In fact, any successful attempt to fit this band using current electric dipole models yielded unphysical parameters suggesting hybrid origin. On the other hand, we found that that band shape may be reproduced by a Weibull profile given by,

$$R(\omega) = R_0 + a \left( \frac{c-1}{c} \right)^{\frac{1-c}{c}} \left[ \left( \frac{\omega - \omega_0}{b} \right) + \left( \frac{c-1}{c} \right)^{\frac{1}{c}} \right]^{c-1}$$

$$e^{\left[ \left( \frac{\omega - \omega_0}{b} \right) + \left( \frac{c-1}{c} \right)^{\frac{1}{c}} \right]^c} + \left( \frac{c-1}{c} \right) \tag{5}$$

where a, b, c are fitting constants with maximum peaking at the experimental $\omega_0$ frequency. Figure 7 shows the temperature dependent fit of peak frequencies ($\omega_0$) that for this case yields $A_{Ph} = 11.21 \pm ^{0.25}_{0.05}$, $T_{Cr} = 78.00 \pm ^{1.3}_{0.6}$ and an exponent $\beta_{Ph} = 0.249 \pm ^{0.002}_{0.006}$ that is a value associated to critical regimes close to temperatures at which a phase transition takes place. In a structural phase transition $\beta \sim 0.25$ points to a delicate balance between a discontinuous first order and a continuous second order. As it was also reported for $BiFeO_3$, [35], this exponent implies a fourth power coefficient zero in the free energy expansion of the order parameter that, in $NdMnO_3$, would be related to magnetoelastic coupling in a fluctuating regime with strong magnon-lattice hybridization. [36]. We name this band as *phonon-like* because its possible closer association with a lattice distortion.

In the case of the band peaking at ~20 cm$^{-1}$ being very sharp (Fig. 6) after a few unsuccessful fit attempts we simply assigned the peak position based on the maximum in the spectrum. The three parameter power fit (Fig. 7) to the temperature dependent peak positions yielded $A_{Sp} = 1.99 \pm ^{0.04}_{0.07}$, $T_{Cr} = 78.00 \pm ^{2.9}_{1.5}$ with power law exponent $\beta_{Sp} = 0.530 \pm ^{0.001}_{0.001}$. This corresponds to the Landau critical exponent for spontaneous magnetization in the ordered phase and accordingly, we call it *spin-like*. It is worth stressing that both mode behavior have the antiferromagnetic transition temperature $T_N \sim 78$ K as the critical temperature $T_{Cr}$.

These mode activity seems analogous to antiferromagnetic resonances earlier reported in the infrared for $YMnO_3$ and $Y_{1-x}Eu_xMnO_3$ by Penney et al [37] and by Goian et al [38] respectively. Penney et al calculated resonant modes for a triangular lattice and concluded that two far infrared excitations were degenerate. However, no data is shown in ref 37 to reach a firm conclusion on the possible relation between their 43 cm$^{-1}$ line and our measurements.

Briefly, our understanding of the room temperature THz feature point to not dismissing the role of the $NdMnO_3$ high temperature orbital disordered phase. At lower temperatures, not being fully dynamical quenched, this results in minute orbital misalignments that translate in randomize non-static eg electrons within orbitals yielding a room temperature collective excitation that condenses at low temperatures into two modes. We foresee that raising orbital coherence in compounds with perosvkite distorted lattices, say; by the absence of Jahn-Teller distortions, would significantly diminish its net electric dipole strength. Those orbital fluctuations may be also be seen as an incipient glass state readily malleable by appropriate doping resulting in enhanced magnetoelectric properties [39, 40]. In doped manganites, it adds to the polaron picture underscoring strong polarizations. [41]

On the other hand, short-range spin correlations in $NdMnO_3$ paramagnetic phase ought to be not totally dismissed. Our reflectivities show at about 120 K slightly higher than $T_N$, the collective band sharpens by a temperature driven mechanism in which Mn moments develop magnetic long-range order assisting electron correlation and localization. This band profile change might also point to spin polarization hybridization with $Mn^{3+}$ ions at the oxygen sites as has been reported for $TbMn_2O_5$ [42].

To our knowledge $DyMnO_3$ is the only member of the $RMnO_3$ (R=Rare Earth) family in which an analogous soft mode has been identified and analyzed in terms of the LST relation and a ferroelectric transition. [43] We hypothesize that the band that has been found to soften and assigned as a magnon is likely related to the hybridized pair reported here.

That electromagnon interpretation, i.e, magnons excited by the ac electric field component of the light, is shared by a number of earlier publications reporting on low frequency infrared absorption resonances in hexagonal $RMnO_3$ (R=Rare Earth). [43-51] Although there is an obvious correlation with our data, transmission techniques for polar or ionic crystals are handicapped by the lack of detection of the macroscopic field associated with longitudinal modes. This is a main argument for measuring near normal reflectivity. A transmission (absorption) spectrum peaking at a transverse optical (TO) mode will yield, at higher frequencies than the TO, a featureless background missing most of the information on longitudinal (and collective) excitations [52] as we have found for temperatures above $T_N$ in $NdMnO_3$ (Fig. 6 a; also [21]). This is also supported by our preliminary results on hexagonal $TmMnO_3$, which show the low frequency collective excitation and magnetoelectric couplings playing a main role.[53] In TmMnO$_3$, our "phonon-like" mode splits at 4 K due to the hexagonal lower symmetry and the "spin-like" soft mode moves toward higher frequencies showing a strong Rare Earth dependence implying a non-negligible role in the spin-phonon coupling as already commented for Fig. 4.

Further understanding on the two soft modes may be found comparing these against available polarized inelastic neutron scattering data for zone center spin wave modes in PrMnO3 [14] and YMnO$_3$ [15]. PrMnO$_3$, NdMnO$_3$ and YMnO$_3$, are expected to have zone center magnons in the same energy scale in spite of having A-type and E-type magnetic ordering being mostly insensitive to details of the crystal structure. [54]. It is also worth mentioning the magnetic order evolution in Nd$_{1-x}$Y$_x$MnO$_3$ within the orthorhombic phase in which by replacing Nd by Y one passes from A-type ordering to pure cycloid at x=0.6 [55]. We also note that YMnO$_3$ may be indexed in the NdMnO$_3$ orthorhombic space group if weak extra X-rays diffraction reflections are assigned to the impurity hexagonal distortion.[4]

Fig. 8 shows inelastic neutron scattering data reported by Pailhès et al [15] for zone center magnons and the A-type PrMnO$_3$ zone center magnon measured by Kajimoto et al [14]. On this graph we superposed our depolarized infrared spectra. The spectral features show an excellent agreement. The infrared maxima and the asymmetry, which is coincident with the energy required to move the spins out of the basal plane (spin gap), may be indicative of dynamic modulation by magnetic moments of structural fluctuations associated with potentially developing a ferroeletric distortion. [56] Spin wave excitations are intertwined with the so-called nuclear collective excitations and they propagate through the crystal as spin waves dressed with atomic fluctuations [15]. The same conclusions may be reached comparing the far infrared spectra of hexagonal TmMnO$_3$ at 4 K against zone center magnons reported by H. J. Lewtas et al for isomorphous hexagonal LuMnO$_3$. [57] That is, our findings matching infrared active and neutron detected magnons suggest that magnons active in the 2 to 9 meV energy range have a direct one to one counterpart in the THz region for all RMnO$_3$ (R=Rare Earth). We then conclude that understanding electromagnons thought as unique transverse feature induced by infrared radiation (e.g.: [44]) ought to be reexamined since infrared detection seem to be consequence of the extraordinary magnetoelectric coupling.

Following this line of thought, and within the framework of the recently proposed generalized LST relation by Resta [18, 58] taking into account the coupling of electric and magnetic fields, the number of soft bands may be considered as a direct consequence of individual magnon coupling. The two hybrid modes may then be though as consequence of coupling magnetoelectrically with individual constants α$_i$ (ω) (i=1,2), [18] conforming quasiparticles assigned to hybrid-electric dipole (lattice)-spin (Goldstone) modes. [59, 60] Their temperature dependence tuned to the developing long-range magnetic order below T$_N$ that, in turn, has been interpreted in frustrated magnets as the onset of induced electric polarization breaking the inversion symmetry. [35, 61, 62]. Basal plane Mn-O distances would optimize below $T_N$ the interplay among charge, spin, orbital, and lattice degrees of freedom. Spin-phonon interactions would involve small fluctuations in ion displacements (say, a perovskite inhibited rotation turning lopsided eg orbitals) maintaining, in the X-ray diffraction metric, lattice symmetry.

Not requiring a specific structural change, and thus, of space group, our results yields a novel view for a second order transition in magnetoelectric compounds. Our spectra are also in consonance with magnetic excitations reported for TbMnO$_3$ in which our highest energy "phonon-like" hybrid soft band correlates to the assigned rotations in the magnetic spiral rotation plane which is expected to couple to the electric polarization yielding hybridization. [63]

## CONCLUSIONS

Summarizing, we discussed temperature dependent NdMnO$_3$ far-infrared reflectivity spectra from 300 K to 4K. Our spectra agree with the expected number of 25 for the space group Pbnm ($D_{2h}^{16}$-Z=4) low temperature phonons. However, phonon band profiles show strong anomalies at the Mn$^{3+}$ and Nd$^{3+}$ magnetic ordering temperatures. No new phonons suggesting a structural phase transition are detected at and below ~ $T_N$.

We assign a collective excitation in the paramagnetic phase at THz energies to $e_g$ electrons in d-orbital fluctuations. It locks at ~$T_N$ into two soft bands undergoing hardening down to 4 K. The phonon-like band centered at ~40 cm$^{-1}$ obeys a power law with critical exponent β~0.25 as for a tri-critical point while for the sharper band at ~20 cm$^{-1}$ the power law is β~0.53 as for Landau magnetic ordering. These bands match zone center polarized inelastic neutron scattering spin wave modes. [14,15] They suggest a second order transition at the critical temperature ~$T_N$, in a temperature driven hybridized spin-lattice (Goldstone) mode picture. The two modes, both from eg electrons in deformed d-orbitals, result from magnetoelectric individual couplings between the magnetic spin wave modes (as measured by inelastic neutron scattering) and the electric dipole collective band, (measured only by reflectivity). It also supports conclusions in hexagonal RMnO$_3$ (R=Rare Earth) by Lee et al [64] suggesting that the magnetoelastic coupling is key in understanding magnetoelectric coupling.

Overall, our measurements give a comprehensive view of the temperature evolution of eg electrons in NdMnO$_3$ entangled d-orbitals helping to associate to observations of electronic induced mechanisms for colossal magnetoresistance or polar ordering in transition metal oxides involving orbital/charge and/or spin fluctuations. [65, 40]

# ACKNOWLEDGEMENTS

NEM is grateful to the CNRS-C.E.M.H.T.I. laboratory and staff in Orléans, France, for research and financial support in performing far infrared measurements. LdelC and NEM thanks the Berliner Elektronenspeicherring - Gesellschaft für Synchrotronstrahlung –BESSYII for financial assistant and beamtime allocation under project 2013_1_120813. NEM also acknowledges partial financial support (PIP 0010) from the Argentinean Research Council (Consejo Nacional de Investigaciones Científicas y Técnicas-CONICET). Funding through Spain Ministry of Science and Innovation (Ministerio de Ciencia e Innovación) under Project MAT2010 Nº -16404 is acknowledged by J AA and MJML

# TABLE I

Dielectric simulation fitting parameters for NdMnO$_3$

| T (K) | $\varepsilon_\infty$ | $\omega_{TO}$ (cm$^{-1}$) | $\Gamma_{TO}$ (cm$^{-1}$) | $\omega_{LO}$ (cm$^{-1}$) | $\Gamma_{LO}$ (cm$^{-1}$) |
|---|---|---|---|---|---|
| 110 | 2.68 | 70.7 | 19.9 | 74.1 | 11.6 |
| | | 76.9 | 20.0 | 78.1 | 45.2 |
| | | 117.7 | 14.5 | 118.9 | 13.4 |
| | | 130.0 | 47.2 | 132.8 | 58.9 |
| | | 174.8 | 9.3 | 178.9 | 5.7 |
| | | 184.5 | 8.4 | 186.2 | 13.8 |
| | | 191.0 | 13.9 | 200.1 | 6.3 |
| | | 202.9 | 12.3 | 207.5 | 10.6 |
| | | 233.6 | 21.2 | 237.5 | 37.4 |
| | | 256.5 | 67.8 | 274.3 | 15.1 |
| | | 277.8 | 9.1 | 280.1 | 8.2 |
| | | 286.8 | 14.8 | 292.8 | 8.8 |
| | | 296.9 | 13.4 | 315.3 | 17.8 |
| | | 314.1 | 96.3 | 316.8 | 21.4 |
| | | 317.0 | 7.0 | 325.0 | 33.5 |
| | | 335.3 | 18.5 | 337.6 | 347.0 |
| | | 390.0 | 20.1 | 391.4 | 25.4 |
| | | 393.5 | 53.7 | 407.7 | 26.0 |
| | | 444.9 | 44.6 | 446.9 | 26.1 |
| | | 467.3 | 23.7 | 468.8 | 9.4 |
| | | 496.3 | 45.0 | 498.3 | 16.0 |
| | | 520.0 | 24.4 | 529.2 | 16.1 |
| | | 557.0 | 16.5 | 559.2 | 23.5 |
| | | 567.0 | 37.9 | 607.1 | 63.8 |
| | | 669.2 | 166.0 | 673.5 | 43.9 |

| T (K) | $\varepsilon_\infty$ | $\omega_{TO}$ (cm$^{-1}$) | $\Gamma_{TO}$ (cm$^{-1}$) | $\omega_{LO}$ (cm$^{-1}$) | $\Gamma_{LO}$ (cm$^{-1}$) |
|---|---|---|---|---|---|
| | | 70.2 | 20.3 | 73.8 | 13.1 |
| | | 77.4 | 21.0 | 79.3 | 44.5 |
| | | 118.9 | 9.6 | 120.4 | 15.9 |
| | | 129.3 | 46.0 | 132.1 | 60.7 |
| | | 174.8 | 8.9 | 178.9 | 5.6 |
| | | 184.5 | 8.1 | 186.2 | 13.4 |
| | | 190.7 | 14.8 | 201.2 | 6.7 |
| | | 206.2 | 11.1 | 209.5 | 9.0 |
| | | 235.3 | 20.4 | 238.8 | 34.5 |
| | | 255.8 | 65.0 | 274.9 | 16.9 |
| | | 280.0 | 8.4 | 281.40 | 6.0 |

| T (K) | $\varepsilon_\infty$ | $\omega_{TO}$ (cm$^{-1}$) | $\Gamma_{TO}$ (cm$^{-1}$) | $\omega_{LO}$ (cm$^{-1}$) | $\Gamma_{LO}$ (cm$^{-1}$) |
|---|---|---|---|---|---|
| 70 | 2.71 | 286.8 | 13.6 | 292.9 | 9.4 |
| | | 298.5 | 13.9 | 313.1 | 14.5 |
| | | 313.0 | 93.9 | 319.4 | 20.8 |
| | | 317.2 | 6.8 | 326.1 | 29.8 |
| | | 335.9 | 15.8 | 337.1 | 359.2 |
| | | 389.1 | 17.5 | 390.5 | 20.7 |
| | | 394.8 | 51.3 | 408.6 | 25.1 |
| | | 448.8 | 55.5 | 450.4 | 26.5 |
| | | 462.2 | 17.2 | 466.6 | 8.9 |
| | | 496.4 | 41.0 | 498.4 | 17.5 |
| | | 523.5 | 22.6 | 529.6 | 12.7 |
| | | 557.0 | 15.2 | 559.2 | 21.5 |
| | | 567.0 | 387.0 | 609.5 | 62.4 |
| | | 676.8 | 169.7 | 678.5 | 47.9 |

| T (K) | $\varepsilon_\infty$ | $\omega_{TO}$ (cm$^{-1}$) | $\Gamma_{TO}$ (cm$^{-1}$) | $\omega_{LO}$ (cm$^{-1}$) | $\Gamma_{LO}$ (cm$^{-1}$) |
|---|---|---|---|---|---|
| 4 | 2.74 | 68.2 | 15.6 | 72.6 | 10.7 |
| | | 77.4 | 17.6 | 79.0 | 39.3 |
| | | 119.6 | 9.6 | 120.7 | 8.6 |
| | | 129.0 | 44.8 | 132.0 | 59.1 |
| | | 174.8 | 8.7 | 178.9 | 5.5 |
| | | 184.5 | 8.1 | 186.1 | 19.2 |
| | | 190.7 | 23.7 | 201.2 | 7.7 |
| | | 209.6 | 10.7 | 213.0 | 8.7 |
| | | 235.3 | 21.4 | 298.0 | 38.0 |
| | | 255.8 | 63.7 | 274.9 | 15.8 |
| | | 280.0 | 7.1 | 281.4 | 4.8 |
| | | 286.8 | 13.9 | 292.9 | 8.7 |
| | | 298.6 | 13.9 | 313.2 | 11.1 |
| | | 313.0 | 63.5 | 319.4 | 18.8 |
| | | 317.2 | 5.5 | 326.1 | 25.0 |
| | | 336.0 | 15.5 | 337.0 | 291.0 |
| | | 388.8 | 15.4 | 389.9 | 16.2 |
| | | 394.6 | 44.4 | 409.1 | 23.6 |
| | | 448.7 | 44.7 | 449.7 | 23.7 |
| | | 462.0 | 17.3 | 466.2 | 8.3 |
| | | 497.4 | 46.6 | 499.4 | 17.5 |
| | | 519.7 | 18.5 | 527.0 | 14.5 |
| | | 556.7 | 14.6 | 559.3 | 21.2 |
| | | 568.2 | 38.7 | 610.6 | 57.5 |
| | | 675.8 | 151.0 | 679.0 | 44.3 |

# FIGURE CAPTIONS

**Figure 1** (color online) Temperature dependent near normal reflectivity of NdMnO$_3$ from 4 K to 300 K. For better viewing the spectra have been vertically shifted by 0.10 relative to each other.

**Figure 2** (color online) NdMnO$_3$ phonon near normal reflectivity at 4 K; experimental: dots, full line: fit. Inset: As measured semilog plot of the complete far infrared near normal reflectivity of NdMnO$_3$ at 4 K.

**Figure 3** (color online) Relative change in lattice mode profiles due to magnetostriction taking place around T$_N$ ~78 K.

**Figure 4** (color online) Antisymmetric stretching vibrational modes showing diminution of its strength when Nd$^{3+}$ orders magnetically in the ~12 K to ~20 K temperature range. For better viewing the spectra have been frequency shifted down by 5.2 cm$^{-1}$ relative each other

**Figure 5** (color online) Symmetric stretching vibrational band (breathing mode) showed in expanded scale. In absence of magnetic couplings, there are no profile modifications suggesting structural changes below 120 K. For better viewing the spectra have been frequency shifted down by 5.2 cm$^{-1}$ relative each other

**Figure 6** (color online) (a) Lowest frequency instability due to d orbital eg fluctuations from 300 K to 90 K in the paramagnetic phase; (b) As measured band condensation into two hybridized at $T_N$~78 K and mode hardening. For a better viewing the spectra have been vertically shifted by 0.10 relative to each other. Inset: representative fit using the Weibull profile in the 40 cm$^{-1}$ mode at 4K (dot: experimental; line: fit)

**Figure 7** (color online) Power law fits for the phonon-like and spin-like modes (see text).

**Figure 8** (color online) Comparison of zone center spin wave modes at 1.5 K (dot-bar) by Pailhès et al [15] and as measured infrared active soft modes at 4 K (dot). Note that the asymmetry on the highest frequency side suggests correlation to the spin gap found by inelastic neutron scattering. For completeness a hybrid (spin and lattice) mode as it appears in neutron data is also plotted (squares [15]). The square on the energy axis at ~2 meV is the zone center magnon of A-type antiferromagnetic PrMnO$_3$ measured by Kajimoto et al [14]

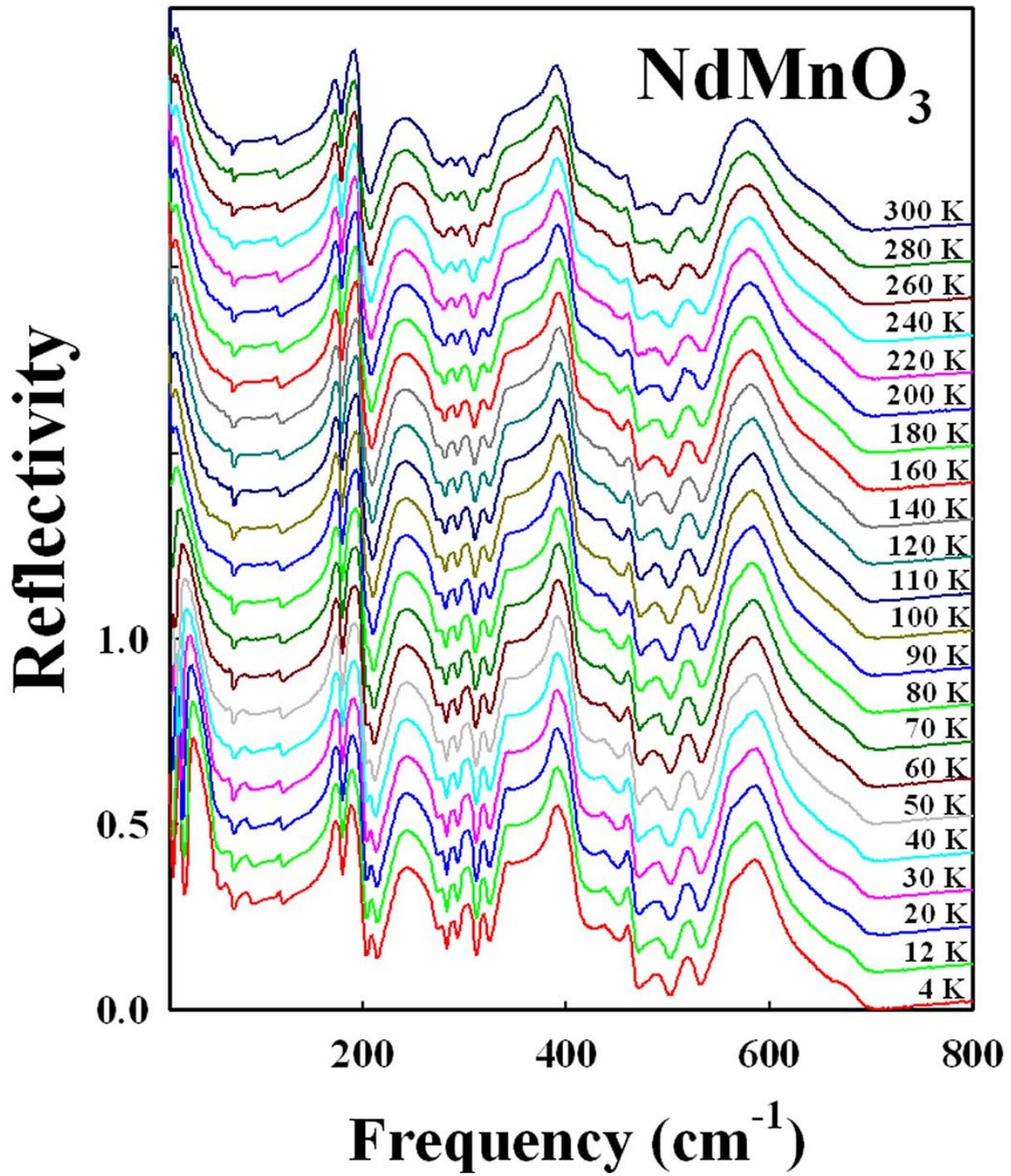

Figure 1
Massa et al

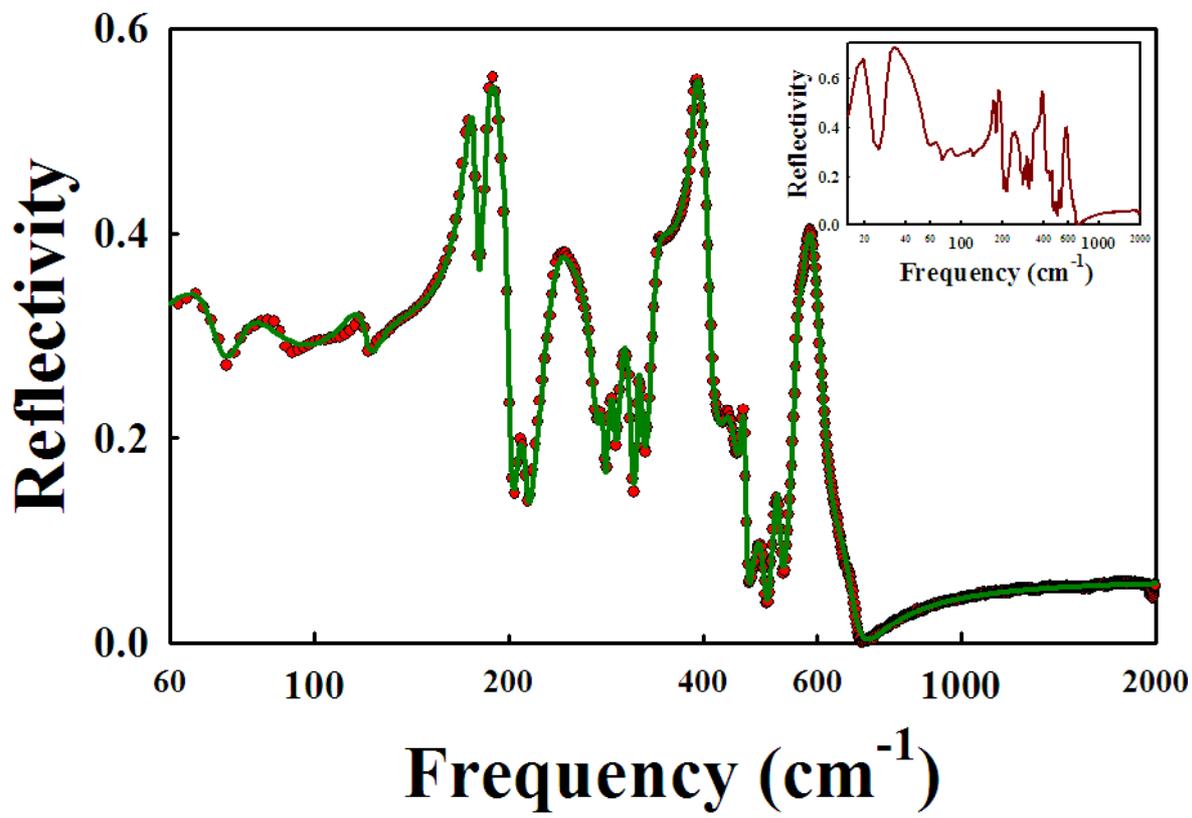

Figure 2
Massa et al

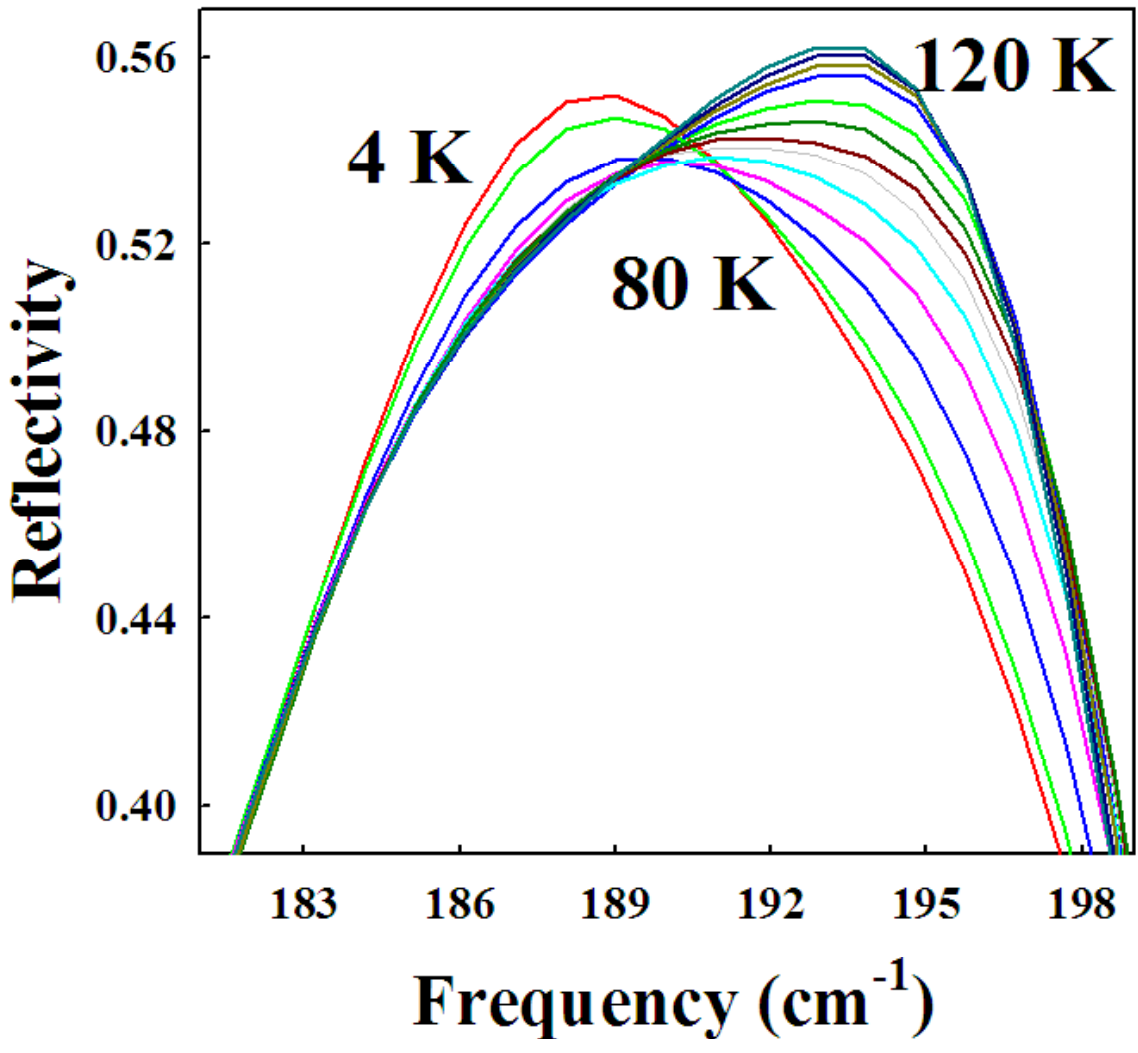

**Figure 3
Massa et al**

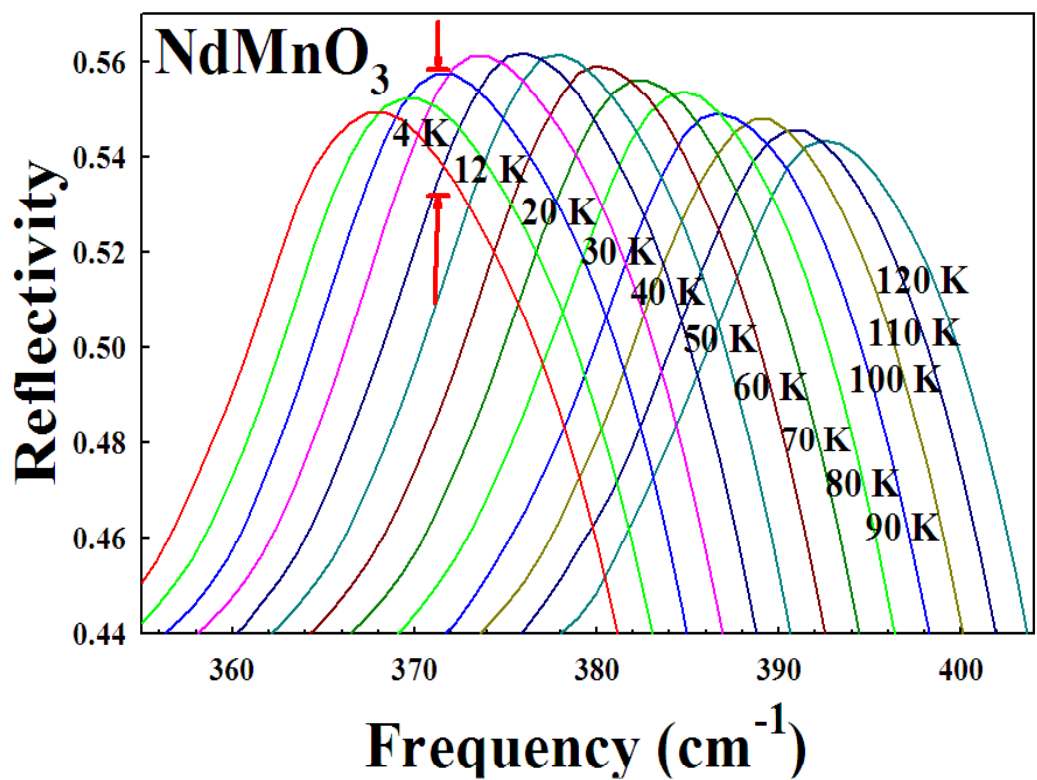

Figure 4
Massa et al

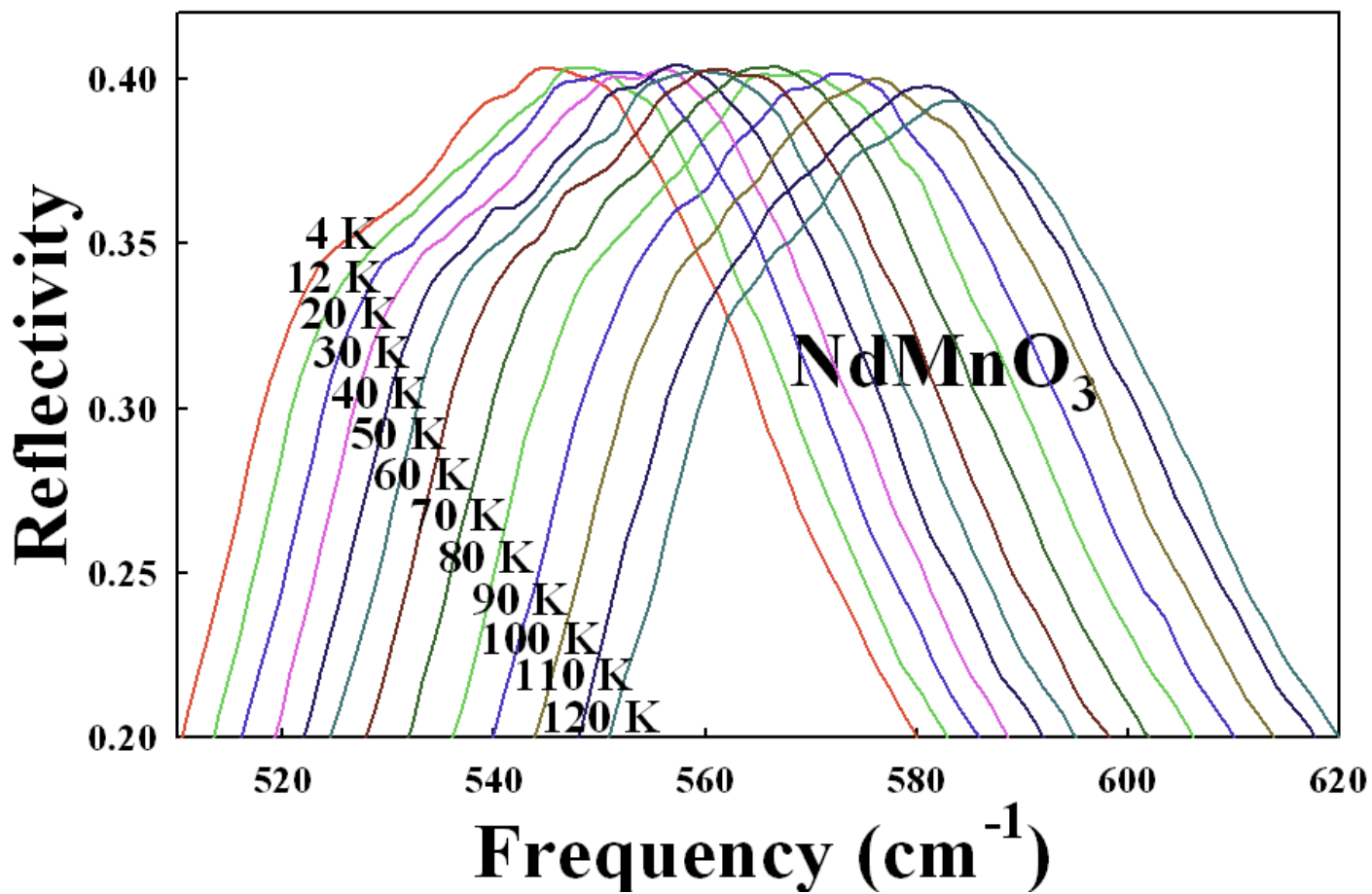

**Figure 5
Massa et al**

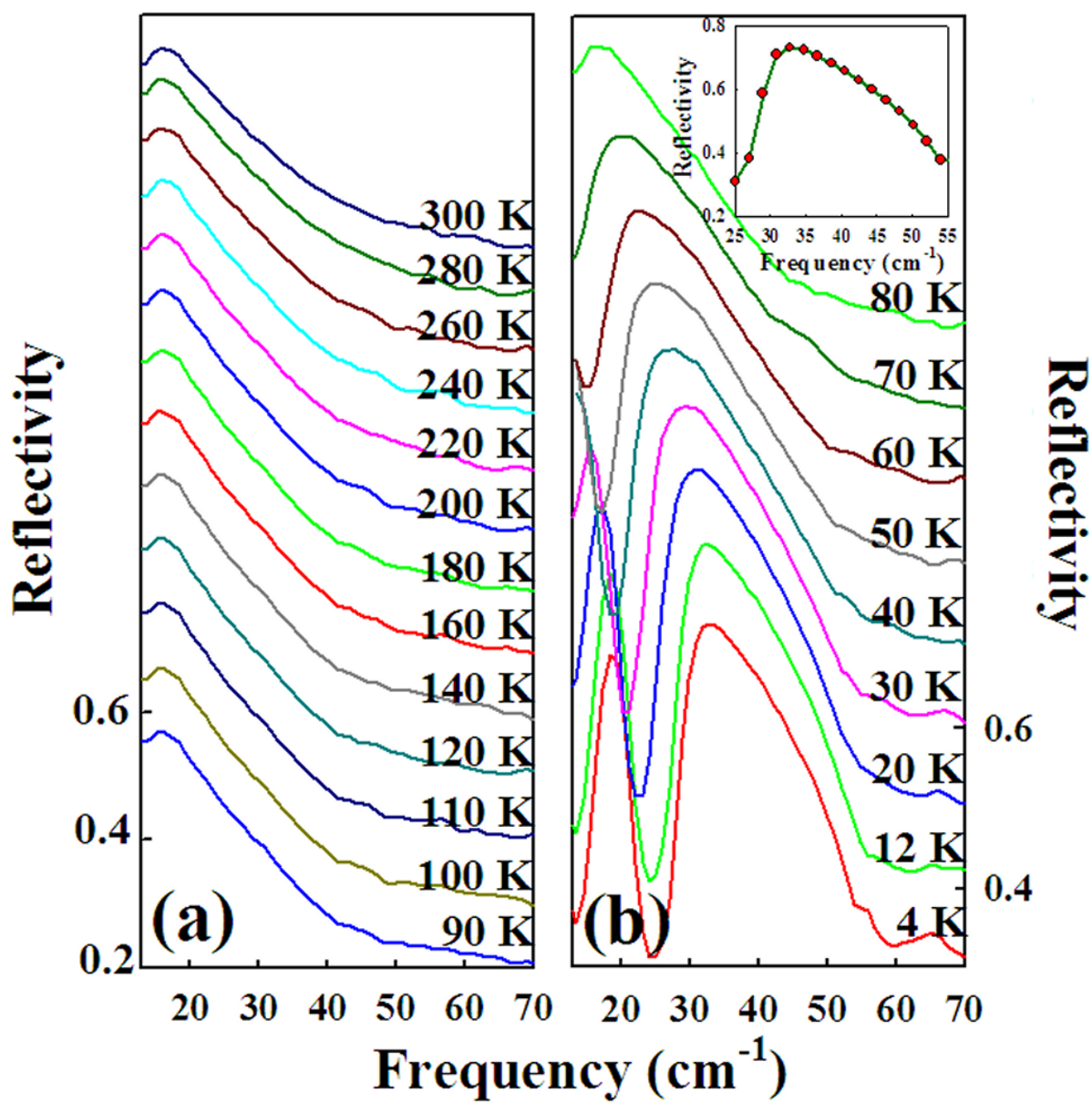

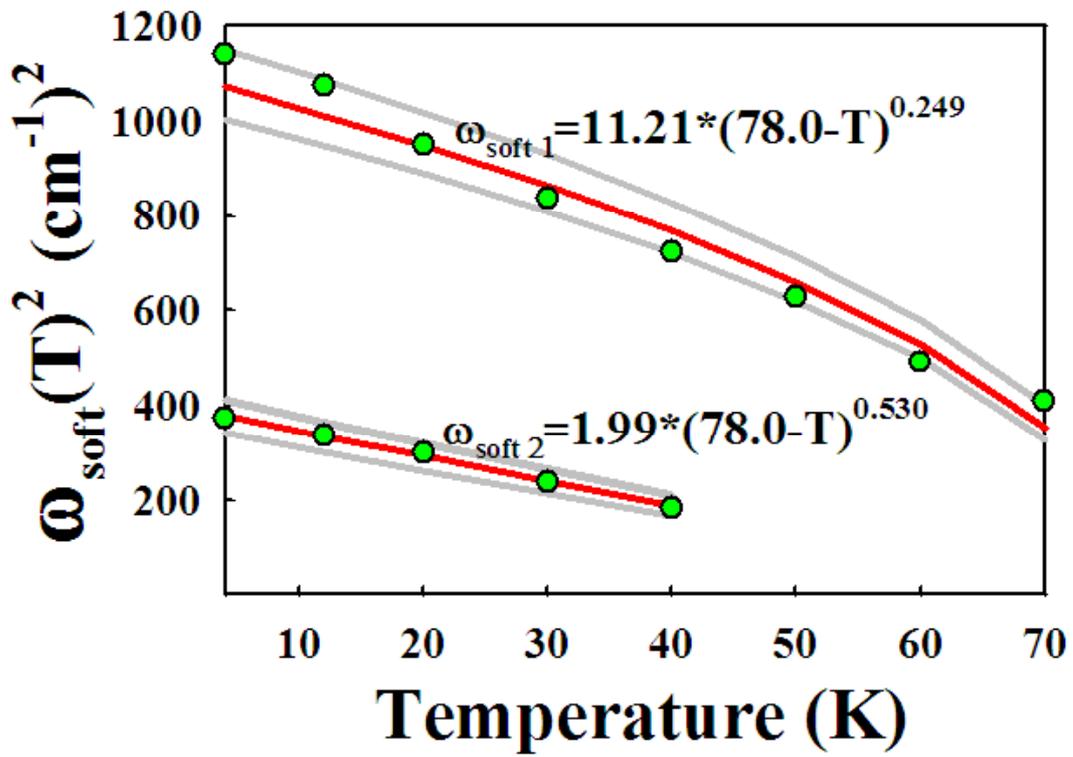

**Figure 7
Massa et al**

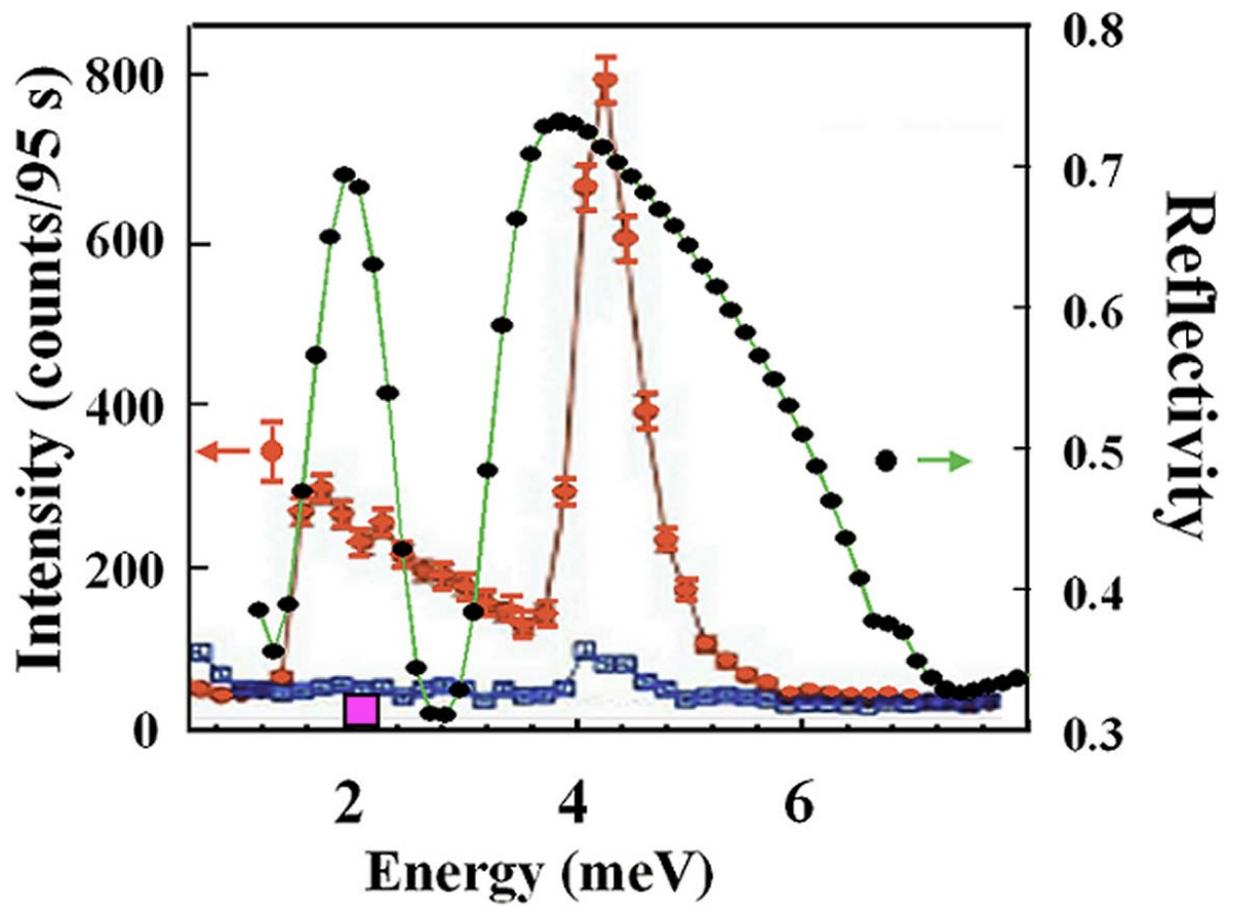

**Figure 8
Massa et al**